\documentclass{mystyle}

\newcommand{\beq}{\begin{equation}}
\newcommand{\eeq}{\end{equation}}
\newcommand{\ba}{\begin{array}}
\newcommand{\ea}{\end{array}}
\newcommand{\beqa}{\begin{eqnarray}}
\newcommand{\eeqa}{\end{eqnarray}}
\newcommand{\lsim}{\stackrel{<}{_\sim}}

\newcommand{\cA}{{\cal A}}
\newcommand{\cO}{{\cal O}}
\newcommand{\dis}{\displaystyle}
\newcommand{\no}{\nonumber}
\newcommand{\op}{Q}

\def\arnps#1#2#3{  {\it Annu. Rev. Nucl. Part. Sci. }{\bf #1}, #3 (#2)}
\def\npb#1#2#3{    {\it Nucl. Phys.}~B {\bf #1}, #3 (#2)}
\def\plb#1#2#3{    {\it Phys. Lett.}~B {\bf #1}, #3 (#2)}

\def\prd#1#2#3{    {\it Phys. Rev.}~D {\bf #1}, #3 (#2)}

\def\prl#1#2#3{    {\it Phys. Rev. Lett. }{\bf #1}, #3 (#2)}
\def\ptp#1#2#3{    {\it Prog. Theor. Phys. }{\bf #1}, #3 (#2)}

\def\zpc#1#2#3{    {\it Z. Phys.}~C {\bf #1}, #3 (#2)}

\def\epjc#1#2#3{   {\it Eur. Phys. J.}~C {\bf #1}, #3 (#2)}
\def\jhep#1#2#3{   {\it JHEP  }{\bf #1}, #3 (#2)}

\def\ibid#1#2#3{{\bf #1}, #3 (#2)}

\begin{document}

\title{Rare decays: theory vs. experiments}

\author{Gino Isidori}

\address{Theory Division, CERN, CH-1211 Geneva 23, Switzerland\\
INFN, Laboratori Nazionali di Frascati, Via E. Fermi 40, I-00044 Frascati, Italy
\\E-mail: Gino.Isidori@cern.ch}

\twocolumn[\maketitle\abstract{
We present an overview of rare $K$, $D$ and $B$ decays.
Particular attention is devoted to those flavour-changing neutral-current 
processes of $K$ and $B$ mesons that offer the possibility of new
significant tests of the Standard Model. The sensitivity of these 
modes to physics beyond the Standard Model and the status of their 
experimental study are also discussed. 

\bigskip

\centerline{CERN-TH/2001-284 $\qquad$ hep-ph/0110255}

}]

\section{Introduction}
Why are we interested in rare decays?
As a general rule, rare processes are particularly 
interesting when their suppression is associated to 
some, hopefully broken, conservation law. 
The most significant examples in this respect are 
proton decay and  $\mu \to e \gamma$: processes 
completely forbidden within the Standard 
Model (SM) that, if observed, would represent an invaluable 
step forward in our understanding of fundamental interactions. 

Conservation laws that so far appear unbroken can also 
be tested by means of heavy mesons. 
However, the most interesting perspectives in 
rare $K$, $D$ and $B$ decays are probably those 
opened  by precision studies of flavour-changing 
neutral currents (FCNCs), or 
transitions of the type
\beq
q_i \to q_j + \left\{ \ba{l}  \nu {\bar \nu} \\ 
\ell^+\ell^- \\ \gamma 
\ea \right.
\eeq
These processes are not completely forbidden 
within the SM, but are generated only at the quantum level 
because of the Glashow--Iliopoulos--Maiani (GIM)
mechanism,\cite{GIM} and are additionally 
suppressed by the hierarchical structure\cite{Wolf}
of the Cabibbo--Kobayashi--Maskawa 
(CKM) matrix.\cite{CKM}  
FCNCs are thus particularly well suited to 
study the dynamics of quark-flavour mixing, within 
and beyond the SM. As a matter of fact, some of these 
processes (such as $K_L \to \mu^+\mu^-$) 
have played an important  role in the historical
formulation of the SM. 

\begin{figure}[t]
\vspace*{-4.0 true cm}
\epsfxsize205pt
\figurebox{205pt}{205pt}{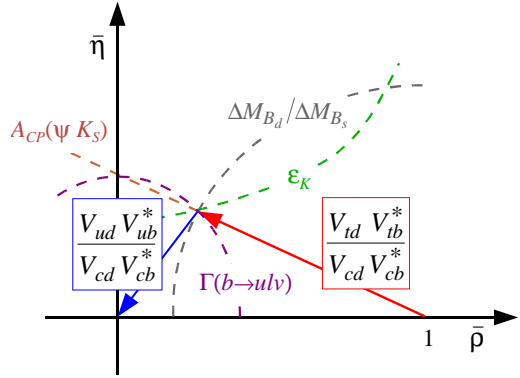}
\vspace*{-1.0 true cm}
\caption{Definition of the reduced CKM unitarity triangle,\protect\cite{Buras_CKM} 
with the indication of the most significant experimental constraints
currently available.}
\label{fig:UT}
\end{figure}

As discussed by many speakers at this conference,  
the CKM mechanism of quark-flavour mixing is 
in good agreement with all data available at present.
The recent measurements
of CP violation in the $B_d$ system\cite{Babar,Belle}  
add a new piece of information that fits 
remarkably within  the overall picture.\cite{Buras2001}
One could therefore doubt about the need for new 
tests of the SM in the sector of (quark) flavour physics. 
However, there are at least two arguments why 
the present status cannot be considered conclusive 
and a deeper study of FCNCs is very useful:
\begin{itemize}
\item
The information used at present to constrain the CKM matrix 
and, in particular, the unitarity triangle,\cite{Buras2001} 
is obtained only from charged currents (i.e.~form tree-level amplitudes)
and $\Delta F=2$ loop-induced processes (see Fig.~\ref{fig:UT}). 
In principle, rare $K$ and $B$ decays mediated by FCNCs  
could also be used to extract indirect 
information on the unitarity triangle. However, 
either because of experimental difficulties or because 
of theoretical problems, the quality of this 
information is very poor at present, with at least $\cO(100\%)$ 
uncertainties. Since new physics could affect in a very different 
way $\Delta F=2$ and $\Delta F=1$ loop-induced amplitudes
[e.g.~with $\cO(100\%)$ effects in the former and  $\cO(10\%)$
in the latter], it is mandatory to improve the quality 
of the FCNC information. 
\item
Most of the observables used in the present fits, 
such as $\epsilon_K$, $\Gamma(b\to u \ell \bar{\nu})$ or $\Delta M_{B_d}$,
suffer from irreducible theoretical errors at the 
10$\%$ level (or above). In the perspective of reaching 
a high degree of precision, it would be desirable to 
base these fits only on observables with theoretical errors
at the percent level (or below), such as the CP asymmetry in $B\to J/\Psi K_S$. 
As we shall see, a few rare $K$ and $B$ decays could offer this 
opportunity.
\end{itemize}

\noindent
Motivated by the above arguments, most of this talk is 
devoted to $K$ and $B$ decays that offer the 
possibility of precision FCNC studies.
In particular, $K\to\pi\nu\bar{\nu}$ decays, 
the so-called {\em golden modes} of $K$ physics, 
are discussed in Section 2; in this section we 
shall also make some general remarks about 
non-standard contributions to FCNCs. 
Rare $K$ decays with a charged lepton pair
in the final state are discussed in Section 3. 
A general discussion about inclusive FCNC $\Delta B=1$ 
transitions is presented in Section 4. Section 5 is devoted to 
$B \to X_{s,d}  \gamma$, whereas inclusive and 
exclusive $B$ decays with a charged lepton pair
in the final state are analysed in Section 6. 
A brief discussion about other processes, including  
$D$ decays and lepton-flavour violating modes is 
presented in Section 7. The overall picture is 
summarized in the Section 8. 

\section{FCNCs in $K$ decays: the golden 
$K\to\pi\nu\bar{\nu}$ modes}

\begin{figure}[t]
\vspace*{ -0.2 true cm} 
\epsfxsize200pt
\epsfysize80pt
\figurebox{200pt}{80pt}{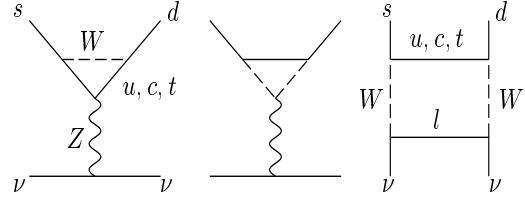}
\caption{One-loop diagrams contributing to
 the $s \to d \nu \bar{\nu}$ transition.}
\label{fig:Kpnn}
\end{figure}

\noindent
The $s \to d \nu \bar{\nu}$ transition is one 
of the rare examples of weak processes whose 
leading contribution starts at $\cO(G^2_F)$. At the one-loop 
level it receives contributions only from $Z$-penguin and 
$W$-box diagrams, as shown in Fig.~\ref{fig:Kpnn}, 
or from pure quantum electroweak effects.
Separating the contributions to the one-loop amplitude according to the 
intermediate up-type quark running inside the loop, we can write
\beqa 
&& \cA(s \to d \nu \bar{\nu}) = \sum_{q=u,c,t} V_{qs}^*V_{qd} \cA_q  \no\\
&& \qquad \sim \left\{ \begin{array}{ll} \cO(\lambda^5
m_t^2)+i\cO(\lambda^5 m_t^2)\    & _{(q=t)}   \\
\cO(\lambda m_c^2 )\ + i\cO(\lambda^5 m_c^2)     & _{(q=c)} \\
\cO(\lambda \Lambda^2_{\rm QCD})     & _{(q=u)}
\end{array} \right. \quad
\label{uno}
\eeqa
where $V_{ij}$ denote the elements of the CKM matrix. 
The hierarchy of these elements 
would favour  up- and charm-quark contributions;
however, the {\em hard} GIM mechanism of the perturbative calculation
implies $\cA_q \sim m^2_q/M_W^2$, leading to a completely 
different scenario. As shown on the r.h.s.~of (\ref{uno}), 
where we have employed the standard CKM phase convention 
($\Im V_{us}=\Im V_{ud}=0$) and 
expanded the $V_{ij}$ in powers of the 
Cabibbo angle ($\lambda=0.22$),\cite{Wolf}
the top-quark contribution dominates both real and
imaginary parts.\footnote{~The $\Lambda^2_{\rm QCD}$ factor in the last
line of (\protect\ref{uno})
follows from a naive estimate of long-distance effects.}
This structure implies several interesting consequences for
$\cA(s \to d \nu \bar{\nu})$:
\begin{enumerate}
\item[a.] it is dominated by short-distance
dynamics, therefore its QCD corrections are
small and calculable in perturbation theory; 
\item[b.] it is very sensitive to $V_{td}$, which 
is one of the less constrained CKM matrix elements;
\item[c.] it is likely to have a large CP-violating phase; 
\item[d.] it is very suppressed within the SM and thus 
very sensitive to possible new sources of quark-flavour mixing.
\end{enumerate}

\noindent 
Short-distance contributions to $\cA(s \to d \nu \bar{\nu})$,
within the SM, can efficiently be described 
by means of a single effective dimension-6 operator:
\beq
Q^{\nu}_{L}= \bar{s}_L\gamma^\mu d_L~\bar{\nu}_L \gamma_\mu \nu_L~.
\label{eq:QnL}
\eeq
Both next-to-leading-order (NLO) QCD corrections\cite{BB,MU,BB2}
and $\cO(G_F^3 m_t^4)$ electroweak corrections\cite{BBmt} to the 
Wilson coefficient of $Q^{\nu}_{L}$ have been calculated,
leading to a very precise description of the partonic amplitude.
In addition, the simple structure of $Q^{\nu}_{L}$ has 
two important advantages: 
\begin{itemize}
\item{} The relation between partonic and hadronic amplitudes 
is quite accurate, since hadronic matrix elements
of the $\bar{s} \gamma^\mu d$ current between a kaon and a pion
are related by isospin symmetry to those entering $K_{l3}$ 
decays, which are experimentally well known. 
\item{} The lepton pair is produced in a state of definite CP 
and angular momentum, implying that the leading SM contribution 
to $K_L \to \pi^0  \nu \bar{\nu}$ is CP-violating.
\end{itemize}

\subsection{SM uncertainties}
The dominant theoretical error in estimating 
the $K^+\to\pi^+ \nu\bar{\nu}$ rate 
is due to the subleading, but non-negligible  
charm contribution. Perturbative NNLO corrections 
in the charm sector have been estimated\cite{BB2} 
to induce an error in the total rate of around  $10\%$, which
can be translated into a $5\%$ error in the determination of 
$|V_{td}|$ from ${\cal B}(K^+\to\pi^+ \nu\bar{\nu})$. 
Recently, also non-perturbative effects introduced 
by the integration over charmed degrees of freedom
have been analysed\cite{Falk_Kp} and turns 
out to be within the error of NNLO terms.
Finally, genuine long-distance effects associated 
to light-quark loops have been shown\cite{LW} 
to be negligible with respect to the uncertainties 
from the charm sector.

The case of $K_L\to\pi^0 \nu\bar{\nu}$ is even cleaner from the
theoretical point of view.\cite{Litt} 
Because of the  CP
structure, only the imaginary parts in (\ref{uno}) 
--where the charm contribution is absolutely negligible--
contribute to $\cA(K_2 \to\pi^0 \nu\bar{\nu})$. Thus 
the dominant direct-CP-violating component 
of $\cA(K_L \to\pi^0 \nu\bar{\nu})$ is completely 
saturated by the top contribution, 
where  QCD corrections 
are suppressed and rapidly convergent. 
Intermediate and long-distance effects in this process
are confined only to the indirect-CP-violating 
contribution\cite{BB3} and to the CP-conserving one,\cite{CPC} 
which are both extremely small.
Taking into account the isospin-breaking corrections to the hadronic
matrix element,\cite{MP} we can write an
expression for the $K_L\to\pi^0 \nu\bar{\nu}$ rate in terms of 
short-distance parameters, namely\cite{BB2,BB3}
\beqa
&& {\cal B}(K_L\to\pi^0 \nu\bar{\nu})_{\rm SM}~=~4.16 \times 10^{-10} \no \\
&& \qquad \times \left[
\frac{\overline{m}_t(m_t) }{ 167~{\rm GeV}} \right]^{2.30} \left[ 
\frac{\Im(V^*_{ts} V_{td})}{ \lambda^5 } \right]^2~, \qquad 
\eeqa
which has a theoretical error below $3\%$.

\begin{figure}[t]
\vspace*{-4.0 true cm} 
\epsfxsize205pt
\figurebox{205pt}{205pt}{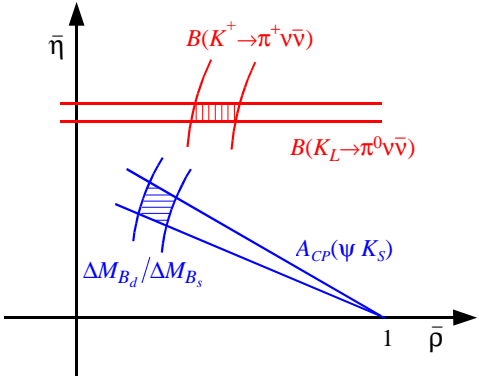}
\vspace*{-1.0 true cm}
\caption{Possible future comparison between 
$K\to \pi\nu\bar\nu$ rates and  
clean $B$-physics observables (in the presence of new physics).\protect\cite{muon}}
\label{fig:gerhard}
\end{figure}

The high accuracy of the theoretical predictions of ${\cal B}(K^+ \to\pi^+
\nu\bar{\nu})$ and ${\cal B}(K_L \to\pi^0 \nu\bar{\nu})$ in terms of modulus
and phase of $\lambda_t=V^*_{ts} V_{td}$ clearly offers
the possibility of very interesting tests of the CKM mechanism. A
measurement of both channels would provide two independent pieces of information on
the unitarity triangle, or a determination of $\bar\rho$ and 
$\bar\eta$.
In principle, as shown in Fig.~\ref{fig:gerhard},
very precise and highly non-trivial tests of the CKM mechanism could
be obtained by the comparison of the following two sets of data:\cite{BB3}
the two $K\to \pi\nu\bar\nu$ rates on one side,  the ratio 
$\Delta M_{B_d}/\Delta M_{B_s}$ and 
the time-dependent CP asymmetry in $B\to J/\Psi K_S$ on the other side.
The two sets are both determined by very different 
loop amplitudes ($\Delta S=1$ FCNCs and $\Delta B=2$ mixing)
and both suffer of very small theoretical errors.

At present the SM predictions of the two  $K\to \pi\nu\bar\nu$ rates
are not extremely precise owing to the limited knowledge of $\lambda_t$. 
Taking into account all the indirect constraints, the allowed range 
is given by\cite{BB2}
\beqa
{\cal B}(K^+ \to\pi^+ \nu\bar{\nu})^{ }_{\rm SM} &=& (0.8 \pm 0.3) 
\times 10^{-10} \qquad  \label{BRK+nnt}
\\
{\cal B}(K_L \to\pi^0 \nu\bar{\nu})^{ }_{\rm SM} &=& (2.8 \pm 1.1) 
\times 10^{-11} \qquad  \label{BRKLnnt}
\eeqa


\subsection{Beyond the SM: general considerations}

As far as we are interested only in rare FCNC 
transitions, we can roughly distinguish 
the extensions of the SM into two big categories: 
\begin{itemize}
\item{\em 
Models with minimal flavour violation}, or 
models where the only source of quark-flavour mixing 
is the CKM matrix (e.g.~the two-Higgs-doublet model 
of type II, the constrained minimal supersymmetric SM,
etc.). In this case non-standard contributions 
are severely limited by the 
constraints from electroweak data. However, we stress 
that the high-precision   obtained by LEP and SLC at the $Z$ peak 
(typically at the per mille level),
refers to observables that receive tree-level contributions 
within the SM. The accuracy on the pure quantum 
electroweak effects barely reaches the 10\% level. 
Thus even within this constrained scenario one can expect 
deviations at the $10\%$--$30\%$ level in observables such as
$K \to\pi \nu\bar\nu$ rates. Detailed calculations 
performed within the flavour-constrained MSSM 
confirm this expectation.\cite{gambino}
In principle these effects could be detected, since they 
are above the intrinsic theoretical errors.
\item{\em 
Models with new sources of quark-flavour mixing},
such as generic SUSY extensions of the SM, models with new 
generations of quarks, etc.
On general grounds this category is the most natural one, 
since we expect some mechanism beyond the SM to be responsible 
for the observed flavour structure. Indeed the case of 
minimal flavour violation can be considered as a particular 
limit of this general category: the limit where the new sources 
of quark-flavour mixing appear only well above the electroweak scale. \\
FCNCs could be dramatically affected by the presence of 
new sources of quark-flavour mixing, if the latter leads to overcoming 
the strong CKM hierarchy. This effect is potentially 
more pronounced in rare FCNC kaon decays, where the CKM
structure implies an $\cO(\lambda^5)$ suppression of the leading amplitude,
than in $B$ decays. This naive expectation
can explicitly be realized in specific and consistent 
frameworks.\cite{Kpnnsusy}$^{-}$\cite{PQS} 
In particular, within the non-constrained MSSM
it is found\cite{BCIRS} that ${\cal B}(K_L\to\pi^0\nu\bar\nu)$ and/or
${\cal B}_{\rm CPV-dir}(K_L\to\pi^0e^+e^-)$ (to be defined later) 
could be enhanced over SM expectations up to one order of magnitude.
\end{itemize}

\noindent
In general it is not easy to compare the sensitivity of different observables
to physics beyond the SM, without making specific assumptions about it,
and at present we have very limited clues about the nature of non-standard 
physics. In this situation, we believe that a useful guiding principle is 
provided by the theoretical cleanliness of a given process. 
In Table~\ref{tab:comp} we compare three well-known examples 
of observables that probe electroweak amplitudes at the quantum level: 
$\Gamma(K_L \to \pi^0 \nu\bar \nu)$, $\Gamma(B \to X_s \gamma)$
(to be discussed later) and the anomalous magnetic moment
of the muon.\cite{g_2} As can be noted, the limited impact of 
QCD effects makes $\Gamma(K_L \to \pi^0 \nu\bar \nu)$ 
a privileged observatory. Of course this comparison 
is a bit provocative and should not be taken too 
seriously (the weak amplitudes probed by the three processes 
are clearly different), but it illustrates well the virtues of 
$K_L \to \pi^0 \nu\bar \nu$.

\begin{table}[t]
\caption{Theoretical cleanliness of $\Gamma(K_L\to \pi^0\nu\bar\nu)$, 
$\Gamma(B \to X_s \gamma)$ and $(g-2)_\mu$: $\delta_{\rm W}$ 
denotes the pure electroweak contribution; $\delta_{\rm QCD}$
the impact of QCD corrections (both perturbative and non-perturbative ones);
$\Delta^{\rm th}$ the overall theoretical uncertainty.}
\label{tab:comp}
\begin{center}
\begin{tabular}{|l|l|l|l|} 
\hline 
\raisebox{0pt}[12pt][6pt]{Observable} & 
\raisebox{0pt}[12pt][6pt]{$\delta_{\rm QCD}/\delta_{\rm W}$} &
\raisebox{0pt}[12pt][6pt]{$\Delta^{\rm th}/\delta_{\rm W}\!\!\!\!\!\! $} \\
\hline
\raisebox{0pt}[12pt][6pt]{$\Gamma(K_L\to \pi^0\nu\bar\nu)$} & 
\raisebox{0pt}[12pt][6pt]{$< 10\%$ } & 
\raisebox{0pt}[12pt][6pt]{$< 3 \%$ } \\
\hline
\raisebox{0pt}[12pt][6pt]{$\Gamma(B\to X_s \gamma)$} & 
\raisebox{0pt}[12pt][6pt]{$\sim 300\%$ } & 
\raisebox{0pt}[12pt][6pt]{$(10$--$15)\%$ } \\
\hline
\raisebox{0pt}[12pt][6pt]{$(g-2)_\mu$ } &  
\raisebox{0pt}[12pt][6pt]{$\sim 4000\%$ } & 
\raisebox{0pt}[12pt][6pt]{$(50$--$100)\%\!\!\!\!\! $ } \\\hline
\end{tabular}
\end{center}
\end{table}

\subsection{Experimental perspectives}
\begin{figure}[t]
\epsfxsize200pt
\epsfysize230pt
\figurebox{200pt}{230pt}{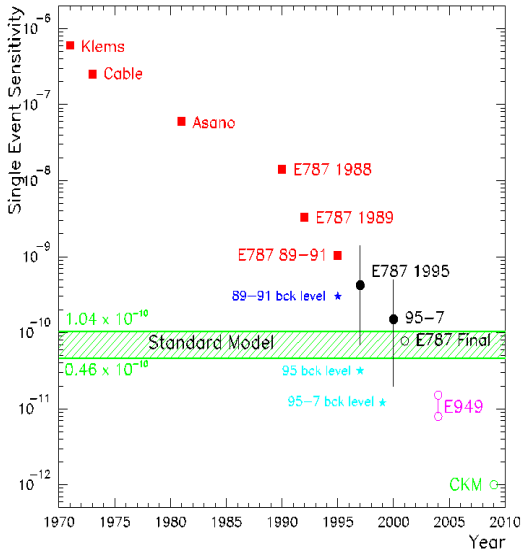}
\vspace*{ -0.4 true cm} 
\caption{History and future prospects in the experimental search 
for $K^+ \rightarrow \pi^+ \nu \bar\nu$.\protect\cite{Litt2000}}
\label{fig:kpnn}
\end{figure}

The search for processes with missing energy and branching ratios 
below $10^{-10}$ is definitely a very difficult challenge, but 
has been proved not to be impossible.\footnote{~Extensive discussions 
about the experimental search for rare $K$ decays can be found 
in the recent literature.\cite{muon,Kettel,Litt2000}}
A strong evidence of $K^+ \to \pi^+ \nu\bar\nu$ has 
been obtained by the E787 experiments at BNL: a single
event was observed in a signal region where the 
background expectation is below 10\%.\cite{E787} 
The branching ratio inferred from this result,
\beq
{\cal B}(K^+ \to\pi^+ \nu\bar{\nu})  = (1.5^{+3.4}_{-1.2}) \times 10^{-10}~,
\label{eq:e787}
\eeq
is consistent with SM expectations, although the error does not allow 
precision tests of the model yet.  E787 has completed its data taking
in 1999 and should soon release a final analysis, including a new sample 
with statistics comparable to all its previous published results. 
In the meanwhile, a substantial 
upgrade of the experimental apparatus has been undertaken, resulting in
a new experiment (BNL-E949) that should start taking data this year,
with the goal of collecting about 10 events (at the SM rate) by 2003.
In the longer term, a high-precision result on this mode will arise 
from the CKM experiment at Fermilab, which aims at a  measurement 
of ${\cal B}(K^+ \to\pi^+ \nu\bar{\nu})$ at the $10\%$ level (see Fig.~\ref{fig:kpnn}).

Unfortunately the progress concerning the neutral mode is much slower.
No dedicated experiment has started yet (contrary to the $K^+$ case) 
and the best direct limit is more than four orders of magnitude above the 
SM expectation.\cite{KTeV_nn} 
An indirect model-independent upper bound  on
$\Gamma(K_L\to\pi^0\nu\bar{\nu})$ can be obtained by the isospin relation\cite{GN} 
\beqa
& \Gamma(K^+\to\pi^+\nu\bar{\nu})~= & \no \\
& \Gamma(K_L\to\pi^0\nu\bar{\nu}) +
\Gamma(K_S\to\pi^0\nu\bar{\nu}) &
\label{Tri}
\eeqa
which is valid for any $s\to d \nu\bar\nu$ local operator of dimension 
$\leq 8$ (up to small isospin-breaking corrections).
Using the BNL-E787 result (\ref{eq:e787}), this implies\cite{Kettel} 
\beq
{\cal B}(K_L\to\pi^0\nu\bar{\nu}) <  2.6 \times 10^{-9}~(90\%~{\rm CL})~.
\label{GNbd}
\eeq
Any experimental information below this figure can be translated into 
a non-trivial constraint on possible new-physics contributions to 
the $s\to d\nu\bar{\nu}$ amplitude. The first experiment that should 
reach this goal is E931a at KEK,\cite{Kettel} at present under construction,
which  will also be the first $K_L\to\pi^0\nu\bar{\nu}$ dedicated experiment. 
The goal of KEK-E931 is to reach a single-event sensitivity (SES) of $10^{-10}$. 
The only approved experiment that could reach the SM sensitivity 
on $K_L\to\pi^0\nu\bar{\nu}$ is KOPIO at BNL,\cite{Kettel} whose goal 
is a SES of $10^{-13}$, or the observation of about 50 signal events (at the SM rate) 
with signal/background $\approx 2$. It is worthwhile to stress that KOPIO will
be rather different from all existing kaon experiments at hadron colliders. 
Using a low-energy micro-bounced beam, KOPIO will be able to measure the 
$K_L$ momentum by means of the time of flight. This measurement, together
with the information on energy and directions of the two $\pi^0$ photons,
substantially enhance the discriminating power against the background
(dominated by $K_L \to 2\pi^0$ with missing photons). 
Unfortunately the construction of KOPIO has not started yet because of 
funding problems; if these can be solved soon, the experiment could 
start to run in 2006. Needless to say that, given the theoretical 
interest and the experimental difficulty, an independent 
experimental set-up dedicated to $K_L\to\pi^0\nu\bar{\nu}$ 
would be very welcome.\cite{muon}

\section{$K\to\pi \ell^+\ell^-$ and  $K\to \ell^+\ell^-$}
Similarly to $K\to\pi\nu\bar{\nu}$, 
short-distance contributions to 
$K\to\pi \ell^+\ell^-$ and  $K\to \ell^+\ell^-$ 
are calculable with high accuracy and 
are highly sensitive to modulus and phase of $\lambda_t$.
However, in these processes the size of long-distance 
contributions is usually much larger because of 
electromagnetic interactions. 
Only in few cases (mainly in CP-violating observables) are
long-distance contributions suppressed and 
is it then possible to extract the interesting short-distance
information. 

\subsection{$K\to \pi \ell^+\ell^-$} 
Contrary to the $s \to \nu\bar\nu$ case, the 
GIM mechanism of the $s \to d \gamma^*$ amplitude 
is only logarithmic.\cite{GilmanW} 
As a result, the  $K \to \pi \gamma^* \to \pi \ell^+\ell^-$ amplitude 
is completely dominated by long-distance dynamics 
and  provides a large contribution to the CP-allowed transitions  
$K^+ \to \pi^+ \ell^+ \ell^-$ and $K_S \to \pi^0 \ell^+ \ell^-$.\cite{EPR}
This amplitude can be described in a model-independent way 
in terms of two form factors, $W_+(z)$ and $W_S(z)$,
defined by\cite{DEIP}
\beqa
& i \dis\int\! d^4x e^{iqx} \langle \pi|T \left\{J^\mu_{\rm em}(x)
{\cal L}_{\Delta S=1}(0) \right\} |
K_i \rangle =& \nonumber \\
&\dis\frac{W_i(z)}{(4\pi)^2}\left[z(p_K+p_\pi)^\mu -(1-r_\pi^2)q^\mu
\right]~,\quad  & \label{eq:tff}
\eeqa
where $q=p_K-p_\pi$, $z=q^2/M_K^2$ and $r_\pi = M_\pi /M_K$.
The two form factors are non singular at $z=0$ and,
because of gauge invariance, vanish to lowest order in 
chiral perturbation theory (CHPT).\cite{EPR}
Beyond lowest order 
two separate contributions to $W_i(z)$ can be identified:
a non-local term, $W_i^{\pi\pi}(z)$, due to the 
$K\to 3\pi\to \pi\gamma^*$ scattering, 
and a local term, $W_i^{\rm pol}(z)$,
which encodes the contributions of unknown 
low-energy constants (to be determined by data).
At $\cO(p^4)$ the local term is simply a constant,
whereas at $\cO(p^6)$ also a linear slope in $z$ arises.
Note that already at  $\cO(p^4)$
chiral symmetry alone does not help to 
relate $W_S$ and $W_+$, or $K_S$ and $K^+$ decays.\cite{EPR}

Recent results on $K^+ \to \pi^+ e^+ e^-$ and
$K^+ \to \pi^+ \mu^+ \mu^-$ by  BNL-E865\cite{E865}
indicates very clearly that, 
owing to a large linear slope, the $\cO(p^4)$ 
expression of $W_+(z)$ is not accurate enough.
This should not be considered as a failure of CHPT, 
rather as an indication that large $\cO(p^6)$ contributions
are present in this channel.
Indeed the $\cO(p^6)$ expression of  
$W_+(z)$ seems to fit the data well. Interestingly, 
this is not only due to a new free parameter
appearing at $\cO(p^6)$, but it is also due to the 
presence of the non-local term. The evidence 
of the latter provides a really  significant test 
of the CHPT approach. 

Knowing $W_+(z)$, we can make reliable
predictions about the  CP-violating 
asymmetry between $K^+ \to \pi^+\ell^+\ell^-$ 
and $K^- \to \pi^-\ell^+\ell^-$  distributions. 
This asymmetry is generated by the interference between 
the absorptive contribution of $W_+(z)$ and the CP-violating 
phase of the $s\to d \ell^+\ell^-$ amplitude, dominated by 
short-distance dynamics.\cite{EPR} The integrated asymmetry 
for $M_{\ell^+\ell^-} \geq 2 M_\pi$ is around $10^{-4}$,  
within the SM, for both electron and muon modes.\cite{DEIP}
A measurement at the $10\%$ level, consistent with zero,
has recently been reported by the HyperCP Collaboration 
at Fermilab.\cite{HyperCP} 
In the near future significant improvements can be expected 
by the charged-kaon extension of the NA48 experiment
at CERN,\cite{NA48Lydia} although the sensitivity is likely to
remain very far from SM expectations.

Similarly to the charged modes, also $K_S \to \pi^0 \ell^+\ell^-$
decays are dominated by long-distance dynamics; however, in this
case non-local terms are very suppressed. To a good approximation,
the $K_S \to \pi^0 e^+e^-$ rate can be written as 
\beq
 {\cal B}(K_S \to \pi^0 e^+ e^-) = 5 \times 10^{-9} \times a_S^2
\eeq
where $a_S$, 
defined by $W^{\rm pol}_S(0)=G_F m_K^2 a_S$, 
is expected to be $\cO(1)$.
The recent bound\cite{NA48KS}  
${\cal B}(K_S \to \pi^0 e^+ e^-) < 1.4 \times 10^{-7}$
is still one order of magnitude 
above the most optimistic expectations, but a measurement or 
a very stringent bound on $|a_S|$ will soon arise from 
the $K_S$-dedicated run of NA48\cite{NA48Lydia} 
and/or from KLOE at Frascati.\cite{KLOE} 

Apart from its intrinsic interest, the determination of 
${\cal B}(K_S \to \pi^0 e^+ e^-)$ has important consequences on
the $K_L \to \pi^0 e^+ e^-$ mode. Here the long-distance part 
of the single-photon exchange amplitude is forbidden 
by CP invariance and the sensitivity to 
short-distance dynamics in enhanced. 
The direct-CP-violating part of the 
$K_L \to \pi^0 \ell^+ \ell^-$ amplitude is conceptually similar 
to the one of $K_L \to \pi^0 \nu \bar{\nu}$: it is calculable 
with high precision, being dominated by the top-quark contribution,\cite{BLMM} 
and is highly sensitive to non-standard dynamics.\cite{BCIRS}
This amplitude interfere with the indirect-CP-violating 
contribution induced by  $K_L$--$K_S$ mixing, leading to\cite{DEIP}
\beqa
&&{\cal B}(K_L \rightarrow \pi^0 e^+ e^-)_{\rm CPV}  =  10^{-12} \no \\
&&\times 
\left[ 15.3 a_S^2 \pm  6.8 \frac{\displaystyle \Im \lambda_t}{\displaystyle 10^{-4}} 
|a_S| + 2.8 \left( \frac{\displaystyle \Im \lambda_t}{\displaystyle 10^{-4}}
\right)^2 \right] \no \\
\label{eq:BKLee}
\eeqa
where the $\pm$ depends on the relative sign between short- and 
long-distance contributions, and cannot be determined in 
a model-independent way. Given the present uncertainty on ${\cal B}(K_S \to
\pi^0 e^+ e^-)$, at the moment we can only set a rough upper limit
of $5.4 \times 10^{-10}$ on the sum of all the CP-violating 
contributions to this mode, to be compared with the direct 
limit of $5.6 \times 10^{-10}$ obtained by KTeV
at Fermilab.\cite{KTeV_ee}

An additional contribution to $K_L \to \pi^0 \ell^+ \ell^-$ 
decays is generated by the CP-conserving long-distance processes
$K_L \to \pi^0 \gamma \gamma \to \pi^0 \ell^+ \ell^-$.\cite{Sehgal}
This amplitude does not interfere with the 
CP-violating one, and recent data\cite{NA48Lydia} on $K_L \to 
\pi^0\gamma\gamma$ (at small dilepton invariant mass)
by NA48 indicate that it is very suppressed, 
with an impact on ${\cal B}(K_L \to \pi^0 e^+e^-)$ at the 
level of few$\times 10^{-12}$ at most. Moreover, if the 
$K_L \to \pi^0 e^+e^-$ were observed, 
the CP-conserving contribution could 
efficiently be isolated by a 
Dalitz plot analysis.

At the moment there exist no definite plans to improve the 
KTeV bound on ${\cal B}(K_L \rightarrow \pi^0 e^+ e^-)$. The future information 
on ${\cal B}(K_S \to \pi^0 e^+ e^-)$ will play a crucial role in this respect: 
if $a_S$ were in the range that maximizes the interference effect in 
(\ref{eq:BKLee}), we believe it would be worths while to start a 
dedicated program to reach sensitivities of $10^{-12}$. 

\subsection{$K_L \to \ell^+ \ell^-$}
\begin{figure}[t]
\vspace*{ -0.2 true cm} 
\epsfxsize120pt
\epsfysize80pt
\figurebox{120pt}{80pt}{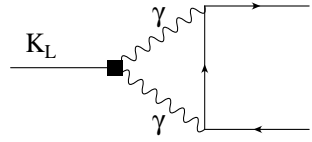}
\vspace*{-0.2 true cm} 
\caption{Two-photon contribution to $K_L \to \ell^+\ell^-$.}
\label{fig:2gamma}
\end{figure}

Both $K_L \to \mu^+ \mu^-$ and  $K_L \to e^+ e^-$ decays are 
dominated by the long-distance amplitude in 
Fig.~\ref{fig:2gamma}. The absorptive part of the 
latter is determined to good accuracy by the two-photon 
discontinuity and is calculable with high precision 
in terms of the $K_L\to \gamma \gamma$ rate. On the other hand, 
the dispersive contribution of the two-photon amplitude 
is a source of considerable theoretical uncertainties. 

In the $K_L\to e^+ e^-$ mode the dispersive integral 
is dominated by a large infrared logarithm 
[$\sim \ln(m_K^2/m^2_e)$], the coupling of which 
can be determined in a model-independent way from  
$\Gamma(K_L\to \gamma \gamma)$. As a result, 
$\Gamma(K_L\to e^+ e^-)$ can be estimated with 
good accuracy\cite{VP} but is almost insensitive to 
short-distance dynamics.

The $K_L\to \mu^+\mu^-$ mode is certainly more interesting from 
the short-distance point of view. Here the two-photon long-distance 
amplitude is not enhanced by large logs and is almost comparable 
in size with the short-distance one,\cite{BB}  
sensitive to $\Re \lambda_t$. Actually short- and long-distance 
dispersive parts cancel each other to a good extent, since the total 
$K_L\to \mu^+\mu^-$ rate (measured with high precision 
by BNL-E871\cite{E871}) is almost saturated by the absorptive 
two-photon contribution:\cite{PDG}
\beqa
{\cal B}(K_L\to\mu^+\mu^-)^{\rm exp} &=& (7.15 \pm 0.16) \times 10^{-9}  \no\\
{\cal B}(K_L\to\mu^+\mu^-)^{\rm abs}_{2 \gamma}  &=&   
 \frac{\alpha_{em}^2 m_\mu^2}{2 m_K^2\beta_\mu } 
 \left[\ln \frac{1-\beta_\mu}{1+\beta_\mu} \right]^2  \no \\ 
\quad \times {\cal B}(K_L \to \gamma\gamma)  &=& (7.00 \pm 0.18) \times 10^{-9}  \no
\eeqa
[$\beta_{\mu} = (1-4 m_\mu^2/m_K^2)^{1/2}$].
The accuracy on which we can bound the two-photon dispersive 
integral determines the accuracy of possible 
bounds on $\Re \lambda_t$. A partial control of the 
$K_L \to \gamma^* \gamma^*$ form factor, which rules the dispersive 
integral, can be obtained by means of $K_L
\to \gamma \ell^+\ell^-$ and $K_L \to e^+e^- \mu^+\mu^-$ spectra; 
additional constraints can also be obtained from model-dependent 
hadronic ansatze and/or perturbative QCD.\cite{BMS,DIP} 
Combining this information, significant 
upper bounds on $\Re \lambda_t$ (or lower bounds on $\bar \rho$) 
have recently been obtained.\cite{E871,KTeV_alpha} 
The reliability of these bounds has still to be fully investigated, 
but some progress can be expected in the near future. 
On the experimental side, a global and model-independent 
analysis could help to clarify the existing discrepancy\cite{KTeV_alpha} 
about the $K_L \to \gamma^* \gamma$ form factor extracted from 
$K_L \to \gamma e^+ e^-$ and  $K_L \to \gamma \mu^+\mu^-$ modes;
a better measurement of the $K_L \to \gamma \gamma$ rate would 
also decrease the overall uncertainty of the absorptive contribution. 
On the theoretical side, the extrapolation of the form factor 
in the high-energy region, which so far requires model-dependent 
assumptions, could possibly be controlled by means of lattice 
calculations.

\begin{figure}[t]
\vspace*{-1.5 true cm} 
\epsfxsize180pt
\figurebox{180pt}{180pt}{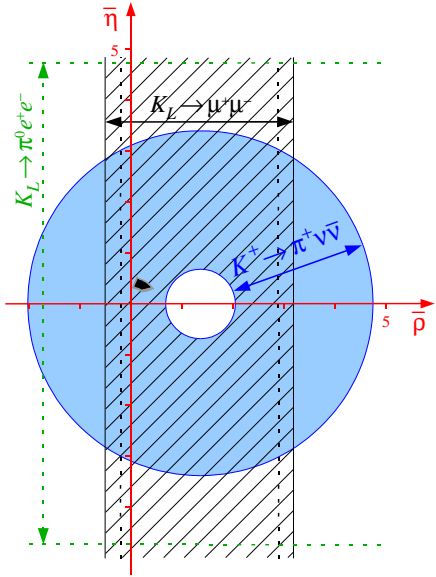}
\vspace*{ 0.5 true cm} 
\caption{Present constraints in the $\bar\rho$--$\bar\eta$ plane from rare $K$ decays
(see text). The small dark region close to the origin denotes the constraints from 
$B$-physics and $\epsilon_K$ (see Fig.~\ref{fig:UT}).}
\label{fig:UT_rare}
\end{figure}

\subsection{Rare $K$ decays and the unitarity triangle}
To conclude the discussion about rare $K$ decays, we summarize 
in Fig.~\ref{fig:UT_rare} the present impact of these modes in 
constraining the $\bar\rho$--$\bar\eta$ plane (complementing 
a recent plot by Littenberg\cite{Litt2000}).
The $K^+\to \pi^+\nu\bar\nu$ constraints are those 
reported by BNL-E787.\cite{Litt2000,E787} 
The bound from $K_{L,S} \to \pi^0 e^+ e^-$
has been obtained by means of Eq.~(\ref{eq:BKLee}), combining the 
recent experimental limits from KTeV\cite{KTeV_ee} and 
NA48\cite{NA48KS} on $K_L$ and $K_S$ decays, respectively. 
Finally, the $K_L \to \mu^+\mu^-$ constraint has been obtained 
by means of the $K_L \to \gamma^* \gamma^*$ form factor 
by D'Ambrosio {\em et al.},\protect\cite{DIP}
combining theoretical and experimental 
errors linearly\cite{Litt2000} (dashed region) 
or in a Gaussian way\cite{KTeV_alpha} (dashed vertical lines).
These bounds are clearly 
less precise than those from $B$-physics; however, the 
comparison is already non-trivial, given the 
different nature of the amplitudes involved. 
Interesting developments in the near future
could arise by an increase of the lower bound 
on ${\cal B}(K^+\to \pi^+\nu\bar\nu)$,
possible if BNL-E787 has collected new signal events.

\section{FCNCs in $B$ decays: generalities}

Inclusive rare $B$ decays such as $B\to X_s\gamma$,
$B\to X_s \ell^+ \ell^-$ and $B\to X_s\nu\bar\nu$
are the natural framework for high-precision
studies of FCNCs in the $\Delta B=1$ sector.\cite{B_rev}  
Perturbative QCD and heavy-quark expansion\cite{Wise} 
form a solid theoretical framework to describe 
these processes: inclusive hadronic rates are related 
to those of free $b$ quarks, calculable in perturbation 
theory, by means of a systematic expansion in 
inverse powers of the $b$-quark mass.

The starting point of the perturbative partonic calculation 
is the determination of a low-energy effective
Hamiltonian, renormalized at a scale $\mu={\cal O}(m_b)$,
obtained by integrating out the heavy degrees 
of freedom of the theory. For $b\to s$ transitions 
--within the SM-- this can be written as
\beq
{\cal H}_{\rm eff}\! =\! - \frac{G_F}{\sqrt{2}}  V_{t s}^\ast  V_{tb}  
\sum_{i=1}^{10,~\nu}  C_i(\mu)  Q_i  + {\rm h.c.} \label{eq:he_DB}
\eeq 
where $\op_{1 \ldots 6}$ are four-quark operators, 
$Q_8$ is the chromomagnetic operator, 
\beqa
\op_7           &=& \dis\frac{e}{4 \pi^2} \bar{s}_L \sigma_{\mu \nu} 
                   m_b b_R F^{\mu \nu}  \\
\op_9           &=& \dis\frac{e^2}{4 \pi^2} \bar{s}_L \gamma^\mu b_L 
                   \bar{\ell} \gamma_\mu \ell  \\
\op_{10}        &=& \dis\frac{e^2}{4 \pi^2} \bar{s}_L \gamma^\mu b_L 
                   \bar{\ell} \gamma_\mu \gamma_5 \ell  
\eeqa
and $Q_\nu$ is the $b\to s$ analogue of $Q_L^\nu$ in Eq.~(\ref{eq:QnL}).
Within the SM, the coefficients of all the FCNC operators 
($Q_7$, $Q_9$, $Q_{10}$ and $Q_\nu$) receive 
a large non-decoupling contribution from top-quark loops
at the electroweak scale. Nonetheless, 
the $m_t$ dependence is not the same for all  
the operators, reflecting a different $SU(2)_L$-breaking structure,
which can be affected in a rather different way by new-physics 
contributions.\cite{radcor}

The calculation of partonic rates then involves three distinct steps: 
i) the determination of the initial conditions of the Wilson coefficients 
at the electroweak scale; ii) the evolution by means of 
renormalization-group equations (RGEs) 
of the $C_i$ down to $\mu={\cal O}(m_b)$; iii) the evaluation of the 
QCD corrections to the matrix elements of the effective operators 
at $\mu={\cal O}(m_b)$. The interesting short-distance dynamics that 
we would like to test enters only in the first step; however, 
the following two steps are fundamental ingredients to reduce and 
control the theoretical error. The status of these steps for the 
three main channels can be summarized as follows:

\begin{description}
\item{$\underline{b \to s \gamma}$} 
As anticipated, QCD corrections play an important role in 
$b \to s \gamma$: the large logarithms generated by 
the mixing of four-quark operators with $Q_7$
(see Fig.~\ref{fig:Q16}) enhance the partonic rate by a factor
of almost three.\cite{Bertolini} Since the mixing of 
$Q_7$ with $Q_{1 \ldots 6}$ vanishes at the one-loop level, 
in this case a full treatment of QCD corrections beyond 
leading logarithms is a rather non-trivial task. This has been 
achieved thanks to the joint effort 
of many authors.\cite{B_rev} In particular, the original calculations of  
SM initial conditions\cite{Adel} and matrix element 
corrections\cite{Greub} have already been confirmed 
by different groups;\cite{Greub_2C,Buras_2C} 
the only part of the SM result 
performed by a single collaboration is the three-loop mixing 
of $Q_7$ and $Q_{1 \ldots 6}$ (notably the most difficult step
of the whole calculation).\cite{Misiak} The accuracy of the
perturbative SM result has also been improved 
with the inclusion of subleading electroweak corrections.\cite{bsgew}
Finally, the initial conditions of Wilson coefficients
have been determined beyond lowest order also 
in the two-Higgs doublet model of type II\cite{Ciuc1} and 
in the constrained MSSM.\cite{bsg_SUSY,bsg_SUSY2}

\begin{figure}[t]
\epsfxsize180pt
\figurebox{180pt}{200pt}{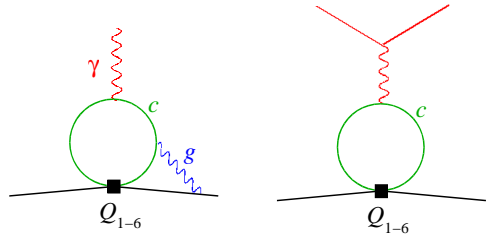}
\vspace*{ -5.0 cm} 
\caption{Representative diagrams for the mixing of 
four-quark operators into $Q_7$ (left) and $Q_9$ (right).}
\label{fig:Q16}
\end{figure}

\item{$\underline{b \to s \ell^+\ell^-}$}
Since $Q_9$ mixes with four-quark operators already 
at the one-loop level (see Fig.~\ref{fig:Q16}), in this case 
QCD corrections are even more important than 
for $b\to s \gamma$.
This fact facilitates the NLO calculation,\cite{bsee_NLO} but enhances
the relative importance of NNLO corrections. 
The latter have recently become available for 
the initial conditions and (part of) the matrix elements.\cite{bsee_NNLO}
\\
Interestingly, the impact of QCD corrections 
is very limited in the axial-current operator $Q_{10}$,
which also contributes to $b \to s \ell^+\ell^-$.
This operator does not mix with four-quark operators 
and is completely dominated by short-distance 
contributions. Together with $Q_\nu$, $Q_{10}$
belongs to the theoretically clean $\cO(G_F^2)$ hard-GIM-protected  
part of the effective Hamiltonian (\ref{eq:he_DB}).
Thus observables more sensitive to $Q_{10}$, 
such as  the forward-backward lepton asymmetry, have a 
reduced QCD uncertainty and a strong sensitivity 
to possible non-standard phenomena.

\item{$\underline{b \to s \nu \bar\nu}$}
QCD corrections to the $b \to s \nu \bar\nu$
amplitude are the same as needed for the $s \to d \nu \bar\nu$ one,\cite{BB,MU,BB2}
with the advantage that charm- and light-quark contributions
are not CKM-enhanced and thus are completely 
negligible also in the real (CP-conserving) part. 
In other words, the only non-trivial step of the 
perturbative calculation for $b \to s \nu \bar\nu$
decays is the determination of the initial condition of $C_\nu$,
which is known with a precision around $1\%$ within the SM.
\\
The experimental upper limit 
\beq
{\cal B}(B \to X_s  \nu\bar\nu) < 6.4 \times 10^{-4}
\eeq
has been announced this year by the 
ALEPH collaboration at LEP.\cite{ALEPH_nn}
This has to be compared with a SM prediction\cite{B_rev} 
of about $3.5 \times 10^{-5}$. Similarly to 
$K\to \pi\nu\bar\nu$ decays, the $b \to s \nu \bar\nu$
transition can probe many new-physics scenarios\cite{GLN} 
and deserve the maximum of attention. Hopefully, the gap between 
SM expectations and experimental limits could decrease
in the next few years at $B$-factory experiments. 
\end{description}

\noindent
The three steps of the perturbative calculation can easily 
be transferred from the $b\to s$ case to the $b \to d$ one,\cite{Greub_bd} 
although the structure of the effective Hamiltonian is richer 
in the latter, owing to the presence of two 
comparable CKM factors ($ V_{t d}^\ast  V_{tb}$ and 
$V_{u s}^\ast  V_{ub}$). Being insensitive to  $V_{td}$, 
$b\to s$ transitions are not interesting for 
precision tests in the $\bar\rho$--$\bar\eta$ plane; 
these processes are particularly useful to constrain 
(or even to detect) possible extensions of the SM.
On the contrary, $b\to d$ transitions are very sensitive to 
$\bar \rho$ and $\bar\eta$, but are clearly disfavoured 
from the experimental point of view because of 
the additional $\cO(\lambda^2)$ suppression.

\section{$B \to X_{s,d} \gamma$}
The inclusive $B \to X_s \gamma$ rate is 
the most significant information that we have
at present on $\Delta B=1$ FCNCs. 
New precise measurements have recently been reported 
by CLEO at Cornell\cite{CLEO_bsg} and 
by BELLE at KEK.\cite{Belle_bsg} 
Combining them with previous determinations,\cite{PDG}
the world average reads
\beq
{\cal B}( B \to X_s \gamma )^{\rm exp} = (3.23 \pm 0.42) \times 10^{-4}
\label{eq:bsg_exp}
\eeq
On the theory side, non-perturbative $1/m_b$ corrections 
are well under control in the total rate. In particular,  
 $\cO(1/m_b)$  corrections vanish in the ratio 
$\Gamma( B \to X_s \gamma )/\Gamma(B \to X_c \ell \nu )$,
and the $\cO(1/m^2_b)$ ones 
are known and amount to few per cent.\cite{Falk}
Also non-perturbative effects associated to charm-quark loops 
have been estimated and found to be very small.\cite{Voloshin,LLW,BIR}
The most serious problem of non-perturbative origin is related 
to the (unavoidable) experimental cut in the photon energy 
spectrum that prevents the measurement from being fully 
inclusive.\cite{LLW,KaganN} With the present cut by CLEO,\cite{CLEO_bsg}
$E_\gamma > 2.0$~GeV, this uncertainty is smaller but non-negligible 
with respect to the error of the perturbative calculation.
The latter is around $10\%$ and its main source is the 
uncertainty in the ratio $m_c/m_b$ that enters ttrhough 
charm-quark loops.\cite{MisiakG} According to a recent analysis 
of all the theoretical uncertainties,\cite{MisiakG}  
the SM expectation is given by
\beq
{\cal B}(B \to X_s \gamma)_{\rm SM} =  (3.73 \pm 0.30) \times 10^{-4}~,
\label{eq:bsg_MG}
\eeq
in good agreement with Eq.~(\ref{eq:bsg_exp}). 
Some comments are in order:
\begin{itemize}
\item{} The central value of the SM prediction in Eq.~(\ref{eq:bsg_MG}) 
is considerably higher than in all previous analyses 
since ${\overline m}_c(\mu)$ has been used, rather than 
the charm pole mass, in the ratio $m_c/m_b^{\rm pole}$
appearing in charm-quark loops. This choice is believed to minimize 
NNLO corrections.\cite{MisiakG}
\item{} The overall scale dependence is very small:
for $\mu \in [m_b/2,2 m_b]$ the central value 
moves by about $1\%$. Therefore the error in Eq.~(\ref{eq:bsg_MG}) is 
an educated guess, whose dominant source is the 
variation of ${\overline m}_c(\mu)/m_b^{\rm pole}$ 
for $\mu \in [m_c , m_b]$ [note that the scale-independent 
ratio ${\overline m}_c(\mu)/{\overline m}_b(\mu)$  is well
within this interval]. Additional uncertainties have 
been combined in quadrature; it is thus more appropriate 
to consider the r.h.s. of Eq.~(\ref{eq:bsg_MG}) as 
central value and standard deviation of a 
Gaussian distribution, rather than as a flat interval. 
\item{} Eq.~(\ref{eq:bsg_MG}) does not include the 
error induced by the extrapolation below the $E_\gamma$ cut. 
This theoretical uncertainty is included in the experimental 
result and, for $E^{\rm min}_\gamma = 2.0$~GeV, is 
around $50\%$ of the error in (\ref{eq:bsg_MG}).
It is worth while to stress that precise data on the 
photon spectrum (above the cut) could help to have 
a better control on this source of uncertainty.\cite{KaganN}
\end{itemize}

\noindent 
The comparison between theory and experiments in 
${\cal B}(B \to X_s \gamma)$ is a great success of the 
SM and has led us to derive many significant bounds on possible 
new-physics scenarios. Non-standard effects of $\cO(1)$ are definitely 
excluded, resulting in stringent constraints of models with 
generic flavour structures, like the unconstrained MSSM.\cite{bsg_nMSSM}
Deviations at the $10\%$--$30\%$ level, as generally 
expected within models with minimal flavour violation,\cite{bsg_SUSY,MisiakG,Barbieri} 
are still possible, and improved measurements of ${\cal B}(B \to X_s \gamma)$ 
are certainly useful to further constrain this possibility.
On the other hand, since the experimental 
error has reached the level of the theoretical one,
it will be very difficult to clearly identify possible 
deviations from the SM, if any, in this observable.

Hopes to detect new-physics signals are still open 
through the CP-violating asymmetry 
\beq
A_{\rm CP}^s = \frac{\Gamma (B \to X_s \gamma ) - \Gamma (\bar{B} \to X_s \gamma ) }
{\Gamma (B \to X_s \gamma ) + \Gamma (\bar{B} \to X_s \gamma ) }~.\
\label{eq:a_s_cp}
\eeq
This is expected to be below $1\%$ within the SM,\cite{Greub_bd,KaganN2} 
but could easily reach $\cO(10\%)$ values beyond the SM, even 
in the absence of large effects in the total $B \to X_s \gamma$
rate. The present measurement of $A_{\rm CP}^s$
is consistent with zero,\cite{bsg_CP} but the  sensitivity
is still one order of magnitude above the SM level.

\medskip

The experimental search for the $B_d \to X_{d} \gamma$
transition is clearly a very hard task. In particular, 
the background generated by $B_d \to X_{s} \gamma$, 
which has a rate 10--20 times larger,\cite{Greub_bd} appears to
be a serious obstacle for the inclusive measurement, 
at least in the short term.
More promising from the experimental point of view are exclusive 
$b\to d$ transitions, such as $B\to \rho \gamma$, which 
have been the subject of recent systematic analyses beyond 
the naive factorization approach.\cite{Pirjol}$^{-}$\cite{Bosch}
%
%
At present the overall theoretical error on 
${\cal B}(B\to \rho \gamma)$ is around $30\%$ and 
is dominated by the uncertainty on the 
hadronic form factors.\cite{Bosch}
Lattice calculations and new experimental data 
could possibly help to reduce this error 
in the near future. 

Similarly to the $b\to s$ case, also in $b\to d$
transitions CP asymmetries are a powerful tool 
to search for new physics. An observable particularly 
appealing both from the experimental and the theoretical 
point of view is the following inclusive asymmetry 
\beq
{\cal B}(B \to X_s \gamma ) + {\cal B}( B \to X_d \gamma ) -\left[ B \to \bar B\right]~.
\label{eq:inc_CPA}
\eeq
Because of the unitarity of the CKM matrix, 
the asymmetry (\ref{eq:inc_CPA}) is expected to be vanishingly small 
within the SM [of $\cO(10^{-9})$]\cite{Soares} and thus 
is an excellent probe of non-standard 
scenarios.

\section{Inclusive and exclusive $b\to s\ell^+\ell^-$ transitions}

The experimental search for FCNC decays of the $B$ meson into 
a charged-lepton pair is just entering into an exciting era:
the first evidence of this type of transition has been announced 
by BELLE at this conference\cite{Belle_rare} 
\beq
{\cal B}(B \to K \ell^{+} \ell^{-}) 
 = (0.75^{+0.25}_{-0.21} \pm 0.09 ) \times 10^{-6}
\eeq
and upper bounds very close to SM expectations have been reported
both by BABAR\cite{Babar_rare} and BELLE\cite{Belle_rare} 
for all the three-body decays of the type 
$B \to (K,K^*) + (\mu^{+} \mu^{-}, e^+ e^-)$.

\begin{figure}[t]
\vspace*{-2.8 cm}
\hspace*{0.2 cm} 
\epsfxsize170pt
\figurebox{170pt}{170pt}{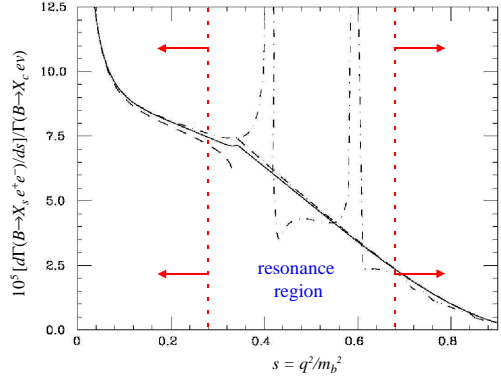}
\vspace*{0.2 cm}
\caption{Dilepton spectrum of the inclusive  
$B \to X_s e^+ e^-$ decay within the SM. The full line denote  
the pure perturbative result (at fixed renormalization 
scale), dashed and dash-dotted lines correspond to estimates 
of non-perturbative $c\bar{c}$ effects.\protect\cite{BIR,ks96}}
\label{fig:bsee}
\end{figure}

Similarly to the $b\to s \gamma$ case, the cleanest
theoretical predictions are obtained for sufficiently 
inclusive observables. Non-perturbative $1/m_b$ 
corrections are well under control in the total rate and 
in the differential dilepton spectrum 
$d(\Gamma \to X_s \ell^+\ell^- )/d s$ 
(but for the end-point $s=M^2_{\ell^+\ell^-}/m^2_b \approx 1 $).\cite{AHHM,BI}
Non-perturbative effects associated to charm-quark loops 
are very large for $M_{\ell^+\ell^-}$ in the region
of the narrow $c\bar c$ resonances (see Fig.~\ref{fig:bsee}),
but they are under control sufficiently far 
from this region.\cite{BIR,ks96} As a result of these two effects, 
the cleanest predictions can be performed for 
$M^2_{\ell^+\ell^-} \lsim 6~{\rm GeV}^2$. 

An important feature of $b\to s\ell^+\ell^+$ transitions
is their sensitivity to the Wilson coefficients $C_{9,10}$.
The latter could be strongly modified 
in several new-physics scenarios, without
observable consequences on $b\to s \gamma$.\cite{AliG}$^{\rm -}$\cite{BHI}
As long as the basis of effective operators is the 
SM one, the purely perturbative dilepton spectrum 
can be written as\cite{bsee_NLO}
\beqa
&& \frac{d}{ds}\Gamma (B \to X_s e^+ e^-) \propto (1-s)^2 \qquad\quad \no \\
&&\quad\ \times \left\{ 4 \frac{s +2}{s}  \left| C_{7} \right|^2 
   + 12 \Re \left[ C^*_7 C_9^{\rm eff}(s)\right] \right. \no\\
&&\qquad\    +\! \left. (1+2s)\left( \left|C_9^{\rm eff}(s) \right|^2 \! + \! 
\left| C_{10} \right|^2 \right)  \right\}~, \qquad
\eeqa
where $C_9^{\rm eff}(s)$ is an appropriate 
combination of $C_9$ and the 
Wilson coefficients of four-quark operators.\cite{bsee_NLO}
At very small $s$, the dominant contribution is that of 
$C_7$, enhanced by $1/s$; however, for $s \approx 0.1$ 
a rapid change of slope is expected because of the interference 
between $C_7$ and $C_9$ (see Fig.~\ref{fig:bsee}).
Since this effect occurs  in the theoretically 
clean part of the spectrum, it could be used to perform new 
high-precision tests of the SM. Even more interesting 
short-distance tests could be performed by means 
of the forward--backward asymmetry of 
the dilepton distribution.\cite{AliG}

\medskip

The high-precision studies allowed by inclusive modes 
will certainly have to wait a few years because of 
experimental difficulties. On the other hand, 
three-body exclusive modes are certainly within 
the reach of $B$-factories. The recent 
bounds\cite{ABHH,BHI} on $B \to (K,K^*) + (\mu^{+} \mu^{-}, e^+ e^-)$
already led us to exclude some of the most 
exotic new-physics scenarios:\cite{LMSS,ABHH,BHI}
non-standard contributions to $C_{9,10}$ can be 
at most of the same order as that of the SM.
Given this situation, it is difficult to detect possible 
deviations from the SM in the total exclusive rates, where 
the theoretical uncertainties are around $30\%$ (or above). 
A much more interesting observable in this respect 
is provided by the lepton forward--backward (FB) asymmetry.
In the $B\to K^* \mu^+\mu^-$ case this is defined as 
\beqa
&& A_{FB}(s)=\frac{1}{d\Gamma(B\to K^* \mu^+\mu^- )/ds}
  \int_{-1}^1 \!\!\! d\cos\theta 
\no\\
&&  
\label{eq:asdef} \qquad
\frac{d^2 \Gamma(B\to K^* \mu^+\mu^- )}{d s~ d\cos\theta}
\mbox{sgn}(\cos\theta)~,
\eeqa
where $\theta$ is the angle between 
$\mu^+$ and $B$  momenta in 
the dilepton centre-of-mass frame. 
Assuming that the leptonic current has only a 
vector ($V$) or axial-vector ($A$) structure, then the
FB asymmetry provides a direct measure of 
the $A$--$V$ interference. Indeed, at LO and 
employing the SM operator basis, one can write
$$
A_{FB}(s) \propto 
  {\rm Re}\left\{  C_{10}^* \left[ s~C_9^{\rm eff} 
    + r(s) \frac{m_b C_7}{m_B}  \right] \right\} 
$$
where $r(s)$ is an appropriate ratio 
of hadronic form factors.\cite{burdman0}
The overall factor ruling the magnitude of $A_{FB}(s)$
is affected by sizeable theoretical 
uncertainties. Nonetheless, there are 
three features of this observable
that provide a clear and independent 
short-distance information:

\begin{figure}[t]
\vspace*{3.0 cm}
\hspace*{-9.5 cm} 
\epsfxsize80pt
\figurebox{80pt}{80pt}{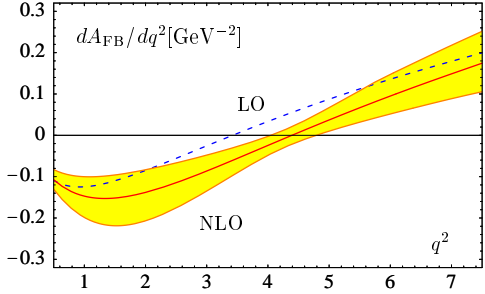}
\vspace*{0.2 cm}
\caption{Forward-backward asymmetry of 
$B^- \to K^{*-} \ell^+\ell^-$ at LO and NLO. 
The band reflects all theo\-retical uncertainties from 
parameters and scale dependence combined.\protect\cite{Martin}}
\label{fig:martin}
\end{figure}

\begin{enumerate}
\item[i.] Within the SM $A_{FB}(s)$ has a zero
in the low-$s$ region (see Fig.~\ref{fig:martin}).\cite{burdman0}
The position of this zero, which depends on the relative 
magnitude and sign of $C_7$ and $C_9$, can be determined 
to a good accuracy within the SM. As recently shown 
by means of a full NLO calculation,\cite{Martin} 
the experimental measurement of $s_0$ could 
allow a determination of $C_7/C_9$ at the  $10\%$ level.
\item[ii.] The sign of $A_{FB}(s)$ around the zero
is fixed unambiguously in terms of the relative sign
of $C_{10}$ and $C_9$:\cite{BHI} within the SM one 
expects $A_{FB}(s) > 0$ for $s > s_0$, for $\bar B$
mesons, as shown in Fig.~\ref{fig:martin}.
\item[iii.] In the limit of CP conservation one expects 
$A^{(\bar B)}_{FB}(s) = - A^{(B)}_{FB}(s)$.
This holds at the per-mille level within the 
SM,\cite{BHI}  where $C_{10}$ has a negligible $CP$-violating phase,
but it could be substantially different in the presence 
of new physics.
\end{enumerate}

\subsection{$B_{s,d} \to \ell^+\ell^-$}
The purely leptonic decays constitute a 
special case among exclusive transitions. Within the SM 
only the axial-current operator, $Q_{10}$, induces a non-vanishing contribution 
to these processes. As a result, the short-distance contribution
is not {\em diluted} by the mixing with four-quark operators.
Moreover, the hadronic matrix element involved is the simplest
we can consider, namely the $B$-meson decay constant
\beq
\langle 0 | \bar q \gamma_\mu \gamma_5 b | \bar B_q (p) \rangle
= i p_\mu f_{B_q} 
\eeq
Reliable estimates of $f_{B_d}$  and $f_{B_s}$ are 
obtained at present from lattice calculations
and in the future it will be possible to cross-check
these results by means of the $ B^+ \to \ell^+ \nu$ rate.
Modulo the determination of $f_{B_q}$, the theoretical
cleanliness of $B_{s,d} \to \ell^+\ell^-$ decays
is comparable to that of $K_L \to \pi^0 \nu \bar{\nu}$ and
$B\to X_{s,d} \nu\bar\nu$.

Compared to their kaon counterparts
($K_L \to \mu^+\mu^-$ and $K_L \to e^+ e^-$)
$B_{s,d} \to \ell^+\ell^-$ decays have the big advantage that
the two-photon amplitude is completely negligible.\footnote{~The smallness of 
the two-photon contribution with respect to the short-distance one
in $B_{s,d} \to \ell^+\ell^-$ decays can easily be deduced from 
a comparison with the $K_L \to \ell^+\ell^-$ case, once 
short- and long-distance contributions 
are rescaled by the appropriate kinematical and CKM factors.}
However, the price to pay is a strong helicity 
suppression for $\ell=\mu$ (and $\ell=e$), or the channels 
with the best experimental signature. Employing the 
full NLO expression\cite{MU,BB2} of $C_{10}$,
we can write 
\beqa
&&{\mathcal B}( B_s \to \mu^+ \mu^-)_{\rm SM} = 3.1 \times 10^{-9} 
\left( \frac{|V_{ts}|}{0.04}  \right)^2 \no\\
&& \times \left( \frac{f_{B_s}}{0.21~\mbox{GeV}} \right)^2 \!\!
\left( \frac{\tau_{B_s}}{1.6~\mbox{ps}} \right)
\left( \frac{ m_t(m_t) }{166~\mbox{GeV} } \right)^{3.12} \quad
\label{eq:BrmmSM} \no \\ \no \\
&&\qquad\quad  \frac{ {\mathcal B}( B_s \to \tau^+ \tau^-)_{\rm SM} }{  
{\mathcal B}( B_s \to \mu^+ \mu^-)_{\rm SM} } = 215~.  \no
\eeqa
The corresponding $B_d$ modes are both 
suppressed by an additional factor $|V_{td}/V_{ts}|^2$ 
$=(4.0 \pm 0.8)\times10^{-2}$. The present experimental 
bound closest to SM expectations is the one obtained by 
CDF, at Fermilab, on $B_{s}\to \mu^+ \mu^-$:
$$
{\mathcal B}(B_{s}\to \mu^+ \mu^-) < 2.6 \times 10^{-6} \quad 
(95 \% \;\;  {\rm CL})~,
$$
which is still very far from the SM level. The latter will 
certainly not be reached before the LHC era.

As emphasized in the recent lit\-era\-ture,\cite{B_mm}$^{-}$\cite{B_mm3}
the purely leptonic decays of $B_s$ and $B_d$
mesons are excellent probes of a specific 
type of new-physics amplitudes, namely 
enhanced scalar (and pseudoscalar) FCNCs. 
Scalar FCNC operators, such as $\bar b_R s_L \bar \mu_R \mu_L$, 
are present within the SM but are absolutely 
negligible because of the smallness 
of down-type Yukawa couplings. On the other hand,
these amplitudes could be non-negligible
in models with an extended Higgs sector.
In particular, within the MSSM, where two Higgs doublets are 
coupled separately to up- and down-type quarks, a strong 
enhancement of scalar FCNCs can occur 
at large $\tan\beta = v_u/v_d$.\cite{B_mm} This effect would 
be practically undetectable in non-helicity-suppressed 
$B$ decays and in $K$ decays (because of the small 
Yukawa couplings), but could enhance $B\to \ell^+\ell^-$
rates by orders of magnitude, up to the  
present experimental bounds.\cite{B_mm2,B_mm3}
The search for these processes is therefore very 
interesting, even if we are still very far from 
the SM level. Experiments at hadron colliders, such as
CDF or, in a long-term perspective, LHCb, are certainly 
advantaged in the search of $B_{s,d} \to \mu^+\mu^-$.
$B$-factory experiments could try to complement 
the picture searching for $B_d \to \tau^+\tau^-$.\cite{B_mm3}

\section{Other rare processes}

\subsection{FCNCs in $D$ decays}
The phenomenology of FCNCs with external up-type quarks, 
such as charm, is completely different from the examples 
discussed so far.\cite{Burdman} In $K$ and $B$ decays 
the short-distance dominance of the clean SM transitions 
is ensured by the presence of the heavy top, which induces non-decoupling 
contributions growing with $m_t$ [as explicitly shown 
in (\ref{uno})]. A similar phenomenon cannot occur in 
$c \to u$ transitions, because of the simultaneous smallness 
of $m_b$ and of the CKM factor $V_{cb} V_{ub}^*$.
Even for $c \to u$ amplitudes with a hard GIM mechanism,
the long-distance contribution dominates within the SM.
As a result, FCNC $D$ decays cannot be used to make 
precision tests of the CKM mechanism.

In cases where it is possible to put a firm upper bound 
on the long-distance contribution, FCNC $D$ decays 
can be used to probe new-physics scenarios.
This possibility has recently been discussed for 
$D \to P \ell^+ \ell^-$, $D \to V \gamma$ and 
$D \to \gamma \gamma$ modes,\cite{Fajfer} where there is still 
a considerable gap between SM expectations 
and experimental limits.
Note that only exotic non-standard scenarios can 
be probed by means of FCNC $D$ decays. 
Indeed, to be clearly identified, the new-physics 
source should produce order-of-magnitude 
enhancements over a long-distance SM amplitude.

\subsection{Lepton-flavour-violating modes}
Decays like $K_L\to \mu e$, $K\to \pi \mu e$ (as well as 
similar $D$ and $B$ modes) are completely forbidden within 
the SM, where lepton flavour is conserved, but are also 
absolutely negligible if we simply extend the model by 
including only Dirac-type neutrino masses.
A positive evidence of any of these processes 
would therefore unambiguously signal new physics, 
calling for non-minimal extensions of the SM.

In exotic scenarios, such as $R$-parity-violating SUSY 
or models with leptoquarks, the $q_i \to q_j \mu e$ amplitude 
can already be generated at tree level. In this case, even 
for high new-physics scales it is possible to generate  
lepton-flavour transitions close to the present experimental limits.
In particular, the bound\cite{E871_me}
\beq
{\cal B}(K_L \to \mu e ) < 4.7 \times 10^{-12}~,
\eeq  
which is the most stringent limit on these types of 
transitions, let us put a bound on leptoquark
masses above $100$~TeV (assuming electroweak
couplings).

In more  conservative scenarios, such as the generic MSSM, 
where $q_i \to q_j \mu e$ transitions occur only at 
the one-loop level and the mechanisms for quark- and 
lepton-flavour mixing are separate, the rates for 
lepton-flavour-violating $K$, $D$ and $B$ decays are naturally 
well below the level of current experimental bounds. 
Nonetheless, as recently shown,\cite{Belaev}  
the branching ratio for $K_L\to \mu e$ in the MSSM with 
generic flavour couplings and $R$-parity conservation 
could be as large as $10^{-15}$, a level that could be 
accessible to a new generation of 
rare-$K$-decay experiments.\cite{muon}
  
\section{Conclusions}
Rare FCNC decays of $K$ and $B$ mesons provide a unique 
opportunity to perform high-precision tests of CP violation 
and flavour mixing, both within and beyond the SM. 

The $B \to X_s \gamma$ rate represents the highest peak 
in our present knowledge of FCNCs: both experimental 
and theoretical errors have reached a comparable level 
of precision, around $10\%$, and the agreement between
theory and data constitutes a highly non-trivial 
constraint for many extensions of the SM. 

The lack of deviations from SM expectations in $\Gamma(B \to X_s \gamma)$
should not discourage the search for other rare FCNC 
observables. As emphasized several times during this talk, 
there are still several observables, such as the forward--backward 
asymmetry in $B \to K^* \ell^+\ell^-$ or the rates of 
$B \to \ell^+ \ell^-$ and $K \to \pi \nu\bar{\nu}$ 
modes, where sizeable deviations from the SM are possible
and are expected in specific new-physics models. 

The measurement of observables with theoretical errors 
at the per-cent level, such as $\Gamma(K_L \to \pi^0 \nu\bar{\nu})$, 
$\Gamma(B \to X_s \nu\bar{\nu})$ and $\Gamma(B_{s,d} \to \mu^+\mu^-)$,
is a very important long-term perspective. Even if new physics 
will first be discovered elsewhere, e.g.~at future hadron colliders, 
the experimental study of such processes would still be 
very useful to investigate the flavour structure of any 
new-physics scenario.

\section*{Acknowledgements}

I am grateful to Juliet and Paolo Franzini for the invitation to this
interesting and very well organized conference. Many thanks are also 
due to Gerhard Buchalla, Andrzej Buras, Gian\-carlo D'Ambrosio, 
Paolo Gambino, Tobias Hurth and Guido Martinelli for several discussions 
on the subject of rare decays and/or comments on the manuscript.
This work is partially supported by the EU-TMR Network EURODAPHNE, ERBFMRXCT-980169.

\end{document}